\documentclass[12pt,epsf]{article}

\newlength{\pubnumber} 

\catcode`\@=11
\@addtoreset{equation}{section}

\def\section{\@startsection{section}{1}{\z@}{3.5ex plus 1ex minus .2ex}
 {2.3ex plus .2ex}{\large\bf}}
\def\subsection{\@startsection{subsection}{2}{\z@}{2.3ex plus .2ex}
 {2.3ex plus .2ex}{\bf}}

\usepackage{graphicx}
\usepackage{dcolumn}
\usepackage{bm}

\usepackage[margin=1in]{geometry}

\usepackage{amsmath}
\usepackage{amsfonts}
\usepackage{amssymb}
\usepackage{hyperref}

\def\beq{\begin{equation}}
\def\eeq{\end{equation}}
\def\beqn{\begin{eqnarray}}
\def\eeqn{\end{eqnarray}}

\def\nolabel{\nonumber }

\def\mod{{\rm mod\ }}

\def\nolabel{\nonumber }

\begin{document}

\begin{titlepage}
\setcounter{page}{1}
\rightline{BU-HEPP-08-03}
\rightline{CASPER-08-05}
\rightline{\tt }

\vspace{.06in}
\begin{center}
{\Large \bf Free Fermionic Heterotic Model Building and Root Systems}
\vspace{.12in}

{\large
        Matthew B. Robinson$^1$ \footnote{m\_robinson@baylor.edu},
        Gerald B. Cleaver$^1$\footnote{gerald\_cleaver@baylor.edu}, and\\
        Markus Hunziker$^2$\footnote{markus\_hunziker@baylor.edu}}
\\
\vspace{.12in}
{\it        $^1$ Center for Astrophysics, Space Physics \& Engineering Research\\
            Department of Physics, One Bear Place \# 97316\\
            Baylor University\\
            Waco, TX 76798-7316\\}
\vspace{.12in}
{\it        $^2$ Department of Mathematics, One Bear Place \# 97328\\
            Baylor University\\
            Waco, TX 76798-7328\\}
\vspace{.06in}
\end{center}

\begin{abstract}
We consider an alternative derivation of the GSO Projection in the free fermionic construction of the weakly coupled heterotic string in terms of root systems, as well as the interpretation of the GSO Projection in this picture.  We then present an algorithm to systematically and efficiently generate input sets (i.e., basic vectors) in order to study Landscape statistics with minimal computational cost.  For example, the improvement at order $6$ is $\approx 10^{-13}$ over a traditional brute force approach, and improvement increases with order.  We then consider an example of statistics on a relatively simple class of models.  
\end{abstract}

\end{titlepage}
\setcounter{footnote}{0}


\section{Free Fermionic Models and the String Landscape}

During the decade following the first string revolution, the focus of string phenomenology was two-fold: 
development of methods for constructing consistent string models with compactified dimensions and searching within the domain of each construction method for those models with at least {\it quasi-realistic} phenomenology \cite{methods}. Many three generation string models were found within a few years. The dominant view back then was that discovery of the true string vacuum (model) was within reach--that the string ``E needle" would eventually be found within the stringstack of but a few trillion vacua. Eventually a handful of quite realistic MSSM or Near-MMSM three generation models were indeed found \cite{mssmbook}, especially following the first {\it Minimal Standard Heterotic String Model} \cite{mshsm}. 

However, following the second string revolution the rise of $M$-theory has taught string phenomenologists the likely impossibility of finding a ``true" string vacuum somewhere on the string/M-landscape composed of {\it at least} $10^{500}$ vacua. In addition to vast number of vacua, all vacua now appear to be on equal footing. Thus, the phenomenological goal has shifted from studying individual string models to better understanding statistically the characteristics of the string/M-models on the landscape, or at least within specific domains \cite{statstudies}. 

The (often overlapping) domains on the landscape frequently correspond to model construction methods. One construction method that has been widely explored in terms of individual models and for which large scale statistical studies are underway is free fermions \cite{fff1,fff2}.
The free fermionic heterotic string has provided many quasi-realistic (Near-)MSSM-like models \cite{nmssm,freegen,optun}, (semi-) GUT models \cite{sguts,psm2}, and GUT models \cite{fsu5}. In the context of the second string revolution, we should now proceed to determine overall pheonenological patterns within these and the many more, as yet undiscovered, free fermionic models.
Therefore, let us review the free fermionic approach as a means of developing a {\it systematic} method of generating models in vast numbers from which the statistics of phenomenological properties may be developed.

\section{The Standard GSO Projection}\label{sec:one}
The first object required to specify a model in the free fermionic heterotic string \cite{fff1,fff2}
is a set
\begin{eqnarray}
A = \{ \vec \alpha^i \in \mathbb{Q}^{64}\cap (-1,1]^{64}| i \in \{1,...,L\in \mathbb{N}\}\}
\end{eqnarray}
where components $\alpha^i_j$, $j=1,\ldots,20$ are boundary conditions for real worldsheet free fermion degrees of freedom of the left-moving supersymmetric string,
 and $\alpha^i_j$ for $j=21,\ldots,64$ are boundary conditions for real worldsheet free fermion degrees of freedom on the right moving bosonic string. In the $\mathbb{C}$ basis (complex fermions), each component of $\vec \alpha^i$ is double counted, and $\vec \alpha^i$ is a $32$ $(10+22)$ component vector (which can be generalized to include left-right paired real fermions).  

The order $N_i$ of a given $\vec \alpha^i$ is defined as\footnote{We include ``0" in the set of natural numbers $\mathbb{N}$ as in set theory, in contrast to its general exclusion in number theory.}
\begin{eqnarray}
N_i \equiv \min \{m\in \mathbb{N} \mid m \alpha^i_j = 0 \; \mod \; 2\;  \forall j\},
\end{eqnarray}
with
\begin{eqnarray}
N_{ij} \equiv {\rm LCM}(N_i,N_j).
\end{eqnarray}
Thus, each component of an $\vec \alpha$ of order $N$ is of the form
\begin{eqnarray}
(\vec \alpha)_i \in {2\mathbb{Z}\over N} \cap [-1,1) \label{eq:elementsofalpha}
\end{eqnarray}

Modular Invariance demands 
\begin{eqnarray}
N_i \; \vec \alpha^i \cdot \vec \alpha^i = 0 \; \mod \; 8 \label{eq:modinv1}
\end{eqnarray}
\begin{eqnarray}
N_{ij} \; \vec \alpha^i \cdot \vec \alpha^j = 0 \; \mod \; 4 \label{eq:modinv2}
\end{eqnarray}
in the $\mathbb{C}$ basis \cite{fff1,fff2}.\footnote{Further, any set of three basis vectors $\alpha^i$, $\alpha^j$, and $\alpha^k$, including cases where $i,\, j,\, {\rm and\\or}\, k$ may be identical. However, this requirement is automatically satisfied by gauge sector basis vectors of the form discussed herein, so this contraint will not be discussed more in this paper.} 
One additional requirement is that each model contain $\vec \alpha^1 = \mathbb{I}$, the 64 real-component vector with every element equal to one.  

In a given free fermionic model, the different sectors of a model are formed by all linear combinations of the $\vec \alpha^i$'s with coefficients $m^k \in \mathbb{N}$ where each coefficient $m^k_i < N_i$.  Each linear combination, or sector, is denoted 
\begin{eqnarray}
\vec V^k = \sum_{i=1}^{L} m^k_i \; \vec \alpha^i \label{eq:defV}
\end{eqnarray}
We can (and will) think of each set of coefficients $m^k$ as an L-dimensional vector in $\mathbb{N}^L$, whose $i^{th}$ component is constrained by the order of $\vec \alpha^i$.

For a given sector $\vec V^k$, a worldsheet fermion $f_j$ transforms as 
\begin{eqnarray}
f_j \rightarrow -e^{i \pi V^k_j}f_j
\end{eqnarray}
around non-contractible loops on the worldsheet.  Thus, for $\mathbb{R}$ fermions, $V^k_j$ must be either 0 or 1, whereas for $\mathbb{C}$ fermions, $V^k_j$ must be rational.  

For each sector, we can form the $U(1)$ charges for the Cartan generators of the unbroken gauge groups (which are in one to one correspondence with the $U(1)$ currents $f^*_jf_j$ for each complex fermion $f_j$);
\begin{eqnarray}
\vec Q_{\vec V^k} \equiv {1 \over 2} \vec V^k + \vec F^k \label{eq:defQ}
\end{eqnarray}
where $\vec F^k$ is a fermion number operator which counts each mode of $f_j$ once and of $f^*_j$ minus once.  Or, in other words, $F^k_i \in \{-1,0,1\}\; \forall\; i$.

The second object required to specify a model in the free fermionic heterotic string is an $L \times L$ matrix $k_{ij}$.  Modular Invariance imposes the following constraints on  $k_{ij}$;
\begin{eqnarray}
k_{ij} + k_{ji} = {1\over2} \vec \alpha^i \cdot \vec \alpha^j \; \mod \; 2 \label{eq:kijone}
\end{eqnarray}
\begin{eqnarray}
k_{ii}+k_{i1} = {1\over4} \vec \alpha^i \cdot \vec \alpha^i - s_i \; \mod \; 2 \label{eq:kijtwo}
\end{eqnarray}
and
\begin{eqnarray}
N_j k_{ij} = 0 \; \mod \; 2
\end{eqnarray}
where $s_i$ is the 4 dimensional spacetime component of $\vec \alpha^i$. 
Furthermore, we demand $k_{ij} \in (-1,1]$.

In this paper, we will concentrate on the (massless) gauge sectors, so we will assume $s_i$=0 and $\alpha^k_i=0$ for $i=0,\ldots,20$ in everything that follows.  Furthermore, we can express the masses of physical states as functions of the charges;
\begin{eqnarray}
\alpha' m^2_{\rm left} = {1\over2} (\vec Q_{\vec V^k,{\rm left}})^2 - {1\over 2}
\end{eqnarray}
\begin{eqnarray}
\alpha' m^2_{\rm right} = {1\over2} (\vec Q_{\vec V^k,{\rm right}})^2 - 1
\end{eqnarray}
We have already demanded that $m_{\rm left}=0$ above.  So, to make $m_{\rm right}=0$, we demand 
\begin{eqnarray}
(\vec Q_{\vec V^k,{\rm left}})^2 = 2
\end{eqnarray}

Now, the GSO projection constraint is;
\begin{eqnarray}
\vec \alpha^i \cdot \vec Q_{\vec V^k} = \sum_{n=1}^L m^k_n k_{in}+s_i \; \mod \; 2 \label{eq:gso}
\end{eqnarray}
with $s_i=0$ in our models.  

\section{Model Building and the Weyl Conditions}

The formulation described in section \ref{sec:one} will produce a set of states which form an abstract root system of simply laced type.  We therefore begin by considering the connections between the approach outlined in section \ref{sec:one} and the Weyl Constraints:  
A set $\Phi$ of vectors $\vec Q^i$ with norm squared 2 form a root system of simply laced type iff they satisfy the following constraints \cite{hum};
\begin{eqnarray}
\vec Q^i \in \Phi &\iff& -\vec Q^i \in \Phi \label{eq:minus}\\
\vec Q^i,\vec Q^j \in \Phi &\Longrightarrow& \vec Q^i - (\vec Q^i \cdot \vec Q^j)\vec Q^j \in \Phi \\
\vec Q^i, \vec Q^j \in \Phi &\Longrightarrow& \vec Q^i \cdot \vec Q^j \in \mathbb{Z} \label{eq:crystal}
\end{eqnarray}
From equations (\ref{eq:defV}) and (\ref{eq:defQ}), it is clear that, given 
\begin{eqnarray}
\vec Q^i = {1\over 2}\sum_{k=1}^{L}m^i_k\vec \alpha^k + \vec F^i
\end{eqnarray}
we can define $\vec Q'^i$ by replacing $\vec F^i$ with $-\vec F^i$ and $m^i_k$ with $N_k - m^i_k$, so that 
\begin{eqnarray}
\vec Q^i + \vec Q'^i = {1\over 2}\sum_{k=1}^{L}m^i_k\vec \alpha^k + \vec F^i + {1\over 2}\sum_{k=1}^{L}(N_k - m^i_k)\vec \alpha^k - \vec F^i = \sum_{k=1}^L N_k\vec \alpha^k \equiv 0 
\end{eqnarray}
So $\vec Q'^i = -\vec Q^i$.  It is clear that $\vec Q'^i$ will satisfy Modular Invariance and the GSO Projection as long as $\vec Q^i$ does, and therefore equation (\ref{eq:minus}) is satisfied.  

Satisfying equation (\ref{eq:crystal}) is a bit trickier.  We can make it more transparent by expanding out only one of the roots, 
\begin{eqnarray}
\vec Q^i \cdot \vec Q^j &=& \bigg({1\over 2}\sum_{k=1}^Lm_k^i \vec \alpha^k + \vec F^i\bigg)\cdot \vec Q^j \in \mathbb{Z} \nolabel \\
&\Rightarrow& {1\over 2}\sum_{k=1}^Lm_k^i \vec \alpha^k \cdot \vec Q^j = a - \vec F^i \cdot \vec Q^j \label{eq:hold1}
\end{eqnarray}
for some $a \in \mathbb{Z}$.  From equations (\ref{eq:defQ}) and (\ref{eq:elementsofalpha}) we can write the general form of $\vec F^i \cdot \vec Q^i$ as 
\begin{eqnarray}
\vec F^i \cdot \vec Q^j &=& {1\over 2} \sum_{k=1}^Lm^j_k \vec \alpha^k\cdot \vec F^i + \vec F^j\cdot \vec F^i = \sum_{k=1}^Lm^j_k{b_{ik}\over N_k} + c
\end{eqnarray}
where $b_{ik},c\in \mathbb{Z}$.  So, (\ref{eq:hold1}) now gives
\begin{eqnarray}
{1\over 2}\sum_{k=1}^Lm_k^i \vec \alpha^k \cdot \vec Q^j = d - \sum_{k=1}^Lm^j_k{b_{ik}\over N_k} 
\end{eqnarray}
where $d$ is now some integer which we can effectually ignore, leaving
\begin{eqnarray}
\sum_{k=1}^Lm^i_k\vec \alpha^k \cdot \vec Q^j = \sum_{k=1}^Lm^j_k {2b_{ik}\over N_k} \label{eq:hold2}
\end{eqnarray}
Without loss of generality, we can take the $m^i_k$'s on the left hand side of (\ref{eq:hold2}) to have a single non-zero unit element.  So, defining $k_{ik} \equiv {2b_{ij} \over N_k}$, we have that in order to satisfy equation (\ref{eq:crystal}), 
\begin{eqnarray}
\vec \alpha^k\cdot \vec Q^j = \sum_{k=1}^Lm^j_k k_{ik}
\end{eqnarray}
where $k_{ik}$ satisfies
\begin{eqnarray}
N_kk_{ik} = 0\;\; \mod \;\; 2,
\end{eqnarray}
which is equivalent to what was stated in section \ref{sec:one}.  In other words, we have derived the general form of the GSO Projection merely by imposing the constraint that our formalism (cf.\ equation (\ref{eq:defQ})) result in an abstract root system of simply laced type.  

\section{Towards Comprehensive Landscape Statistics}

A difficulty in collecting comprehensive statistics in the free fermionic approach (in addition to the sheer size of this domain of the landscape) is in finding an efficient means of systematically generating sets of $\alpha$'s obeying equations (\ref{eq:modinv1}) and (\ref{eq:modinv2}). To date, searches within the landscape of free fermionic heterotic models have been, although wide-ranging, have used {random sampling} \cite{dienes1}. One difficulty with regard to random sampling within the free fermionic portion of the landsacape, is the issue of floating correlations, as discussed in \cite{dienes2}. This problem was shown in \cite{dienes2} to be endemic to random statistical landscape searches and reflects the fact that not all physically distinct string models are equally likely to be sampled in any random search through the landscape or within a specific subspace. This can result in statistical correlations of phenomenological properties of models that ``float'' as a function of sample size. 

While several possible methods were proposed in \cite{dienes2} to overcome this problem, an alternative is to devise a search method that is not random. The remainder of this paper is an exposition of a method of doing exactly this.  As the simplest example, we will consider herein the set of Layer 1, Order 2 calculations. Since the primary purpose of this paper is to outline our general method, a vastly more detailed {\it systematic} search, based on the approach introduced herein, will be the subject of a paper to soon follow \cite{robinson2}.  

Not counting the $\mathbb{I}$ vector, if we have a Layer $L$ set $\{\vec \alpha^i\}$, $i=1,\ldots,L$, with Orders $N_1,N_2,\ldots,N_L$, a brute force approach to finding all Modular Invariant sets requires calculating dot-products for $\big(\prod_{i=1}^LN_i\big)^{22}$ candidate vectors, demanding a total of 
\begin{eqnarray}
L!\bigg(\prod_{i=1}^LN_i\bigg)^{22} \label{eq:original}
\end{eqnarray}
calculations, keeping track of valid sets along the way.  This obviously becomes intractable even for very low Orders and Layers.  Layer one and Order two (which we will denote $L1O2$) will demand $\approx 4\times 10^{6}$, while $L1O4$ demands $\approx 2\times 10^{13}$.  This can be simplified by putting the elements in order in $\vec \alpha^1$.  Then, under each block of similar elements in $\vec \alpha^2$, separately put each element of $\vec \alpha^2$ in order.  This can be repeated for each successive $\vec \alpha$.  While this is  extremely helpful for small $L$, the need to order elements separately for each block of elements above quickly reduces the usefulness of this for higher Layer.  

We can greatly simplify this problem by breaking the set of $\vec \alpha$'s into a tensor product of points in a lower dimensional space.  We elaborate through examples.  For an $\vec \alpha$ of, say, order $2$, the possible elements are $1$ and $0$.  We therefore specify a given order $2$ $\vec \alpha$ by a single number $n_1$, which specifies the number of $1$'s.  The $1$'s are assumed to be moved to the left, so that the $22-n_2$ $0$'s are to the right.  For example, $n_1=4$ would specify the vector 
\begin{eqnarray}
\vec \alpha = (1,1,1,1,0,0,0,0,0,0,0,0,0,0,0,0,0,0,0,0,0,0)
\end{eqnarray}
The dot-product is easy to calculate; $\vec \alpha^2 = n_1$.  

For order $3$, the possible elements are $\pm 2/3,0$.  So, we specify an order $3$ $\vec \alpha$ with two numbers; $n_1$ corresponding to $2/3$, and $n_2$ corresponding to $-2/3$.  In this case, $\vec \alpha^2 = {4\over 9}(n_1+n_2)$.  

This generalizes in the obvious way to higher order.  Higher Layer, however, is a bit more subtle.  Consider $L2O23$ (Layer 2, where the first $\vec \alpha$ is order 2 and the second is order 3), the possible columns when the $\vec \alpha$'s are placed one above the other are 
\begin{eqnarray}
\begin{pmatrix}
1 \\ 2/3
\end{pmatrix}^{n_1},
\begin{pmatrix}
1 \\ -2/3
\end{pmatrix}^{n_2},
\begin{pmatrix}
1 \\ 0
\end{pmatrix}^{n_3},
\begin{pmatrix}
0 \\ 2/3
\end{pmatrix}^{n_4},
\begin{pmatrix}
0 \\ -2/3
\end{pmatrix}^{n_5} \label{eq:nexample}
\end{eqnarray}
So, the dot products in this case will be 
\begin{eqnarray}
(\vec \alpha^1)^2 &=& n_1+n_2+n_3 \nolabel \\
(\vec \alpha^2)^2 &=& {4 \over 9}(n_1+n_2+n_4+n_5) \nolabel \\
\vec \alpha^1 \cdot \vec \alpha^2 &=& {2\over 3}(n_1-n_2)
\end{eqnarray}

For a Layer $L$ set with orders $N_i$, there will be $\prod_{i=1}^L (N_i )- 1$ different $n_i$'s.  Thus, since we know that 
\begin{eqnarray}
\sum_{j=1}^{\prod_{i=1}^L(N_i)-1}n_j \leq 22
\end{eqnarray}
and, defining
\begin{eqnarray}
A \equiv \prod_{i=1}^L (N_i) - 1,
\end{eqnarray}
we find that the total number of calculation which must be performed in this approach is
\begin{eqnarray}
{L! \over A!} \prod_{j=1}^A (22+j). \label{eq:new}
\end{eqnarray}
The improvement of this approach is therefore
\begin{eqnarray}
{\prod_{j=1}^A(22+j) \over A!(A+1)^{22}}. \label{eq:improve}
\end{eqnarray}
The improvement depends on the total Order, not on the Layer.  To see this more clearly, we provide a graph of equation (\ref{eq:improve}) through $A=500$.  

\begin{center}
\includegraphics[scale=1]{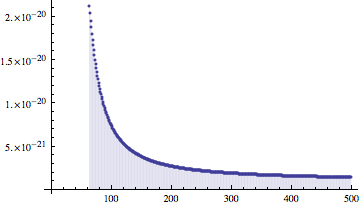}
\end{center}

The improvement is greater for higher order.  This is good news because there are no ordering conventions which help simplify in the original approach (equation \ref{eq:original}) for large Layer ($\approx L\geq 4$), which is where larger orders typically come from. Recall that at $L1O4$, equation (\ref{eq:original}) demanded $\approx 2\times 10^{13}$ calculations.  Equation (\ref{eq:new}), however, demands only $2,300$.
Furthermore, because we can now express any and all dot products as simple linear combinations of the $n_i$'s with rational coefficients, each calculations is straightforward addition, rather than a dot-product of a 22 component vector.  

We now show how to further reduce the complexity of the problem by simplifying the imposition of Modular Invariance.  To begin, in writing out the content of each $n_i$ (as in equation (\ref{eq:nexample})), we will always put the elements in order of decreasing magnitude, starting with the positive elements and then negative.  For example, $L1O6$ would be written as 
\begin{eqnarray}
\bigg({3 \over 3}\bigg)^{n_1},\bigg({2 \over 3}\bigg)^{n_2},\bigg({1\over 3}\bigg)^{n_3},\bigg(-{2\over 3}\bigg)^{n_4},\bigg(-{1\over 3}\bigg)^{n_5}. \label{eq:example2}
\end{eqnarray}
We then write $\vec \alpha^2 = n_1+{4\over 9}n_2+{1\over 9}n_3+{4 \over 9}n_4+{1\over 9}n_5$.  So, Modular Invariance (\ref{eq:modinv1}) demands 
\begin{eqnarray}
6 \vec \alpha^2 &=& 6\bigg(n_1+{4\over 9}n_2+{1\over 9}n_3+{4 \over 9}n_4+{1\over 9}n_5\bigg) = 0 \; \mod \; 8 \nolabel \\
&\Rightarrow& 9n_1+4n_2+n_3+4n_4+n_5 = 0 \; \mod \; 12.
\end{eqnarray}
This is a simple linear equation which can be imposed easily at each step.  However, we can impose this \it a priori \rm as follows.  For an order $N$ $\vec \alpha$ (we assume $L=1$ for now) with the proper ordering (as in (\ref{eq:example2})), the elements of $\alpha$ will come from equation (\ref{eq:elementsofalpha}).  If we demand that $n_A$ satisfy
\begin{eqnarray}
n_A \in -\sum_{j=1}^{A}a_j^2n_j + 2N\mathbb{Z}, \label{eq:squaredconstraint}
\end{eqnarray}
where $a_j$ is $\in \mathbb{Z}$ and specifies the element of $\vec \alpha$ corresponding to $n_j$ according to (\ref{eq:elementsofalpha}).  In other words, $a_j$ is an integer such that ${2a_j \over N} \in [-1,1)$ is the element of $\vec \alpha$ corresponding to $n_j$.  For example, in (\ref{eq:example2}), 
\begin{eqnarray}
a_1 = 3, \; a_2 = 2, \; a_3 = 1, \; etc.
\end{eqnarray}
Our ordering convention guarantees that this will always be an integer.  

For $L\neq 1$, we can impose this constraint on each $\vec \alpha$ separately by demanding that the $n_i$ corresponding to the last non-zero element of each $\vec \alpha$ satisfy this constraint, with the sum going from 1 to the previous $n_i$.  Furthermore, our ordering of the $n_i$'s ensures that an $n_i$ corresponding to a final non-zero element will be unique. Demanding this for each $n_i$ corresponding to the last non-zero element in each $\vec \alpha$ will ensure that (\ref{eq:modinv1}) is always satisfied for each $\vec \alpha$ in the set.  

We can do something similar to impose (\ref{eq:modinv2}).  For the dot-product between $\vec \alpha^i$ and $\vec \alpha^j$, let $n_k$ correspond to the last element which contributes to $\vec \alpha^i \cdot \vec \alpha^j$.  For example, in (\ref{eq:nexample}) this would be $n_2$.  Once again, our ordering convention for the $n_i$'s ensures that each will be unique.  If we impose
\begin{eqnarray}
n_k \in -\sum_{j=1}^{k-1} a_jb_j n_j + {N_iN_j \over N_{ij}} \mathbb{Z}, \label{eq:dotprodcon}
\end{eqnarray}
where $a_j$ is the integer defining an element of $\vec \alpha^i$ and $b_j$ is the integer defining an element of $\vec \alpha^j$, then (\ref{eq:modinv2}) will be satisfied. 

To see this, consider $L3O223$ as an example.  We will have
\begin{eqnarray}
& &\begin{pmatrix}
1 \\ 1 \\ 2/3
\end{pmatrix}^{n_1}
\begin{pmatrix}
1 \\ 1 \\ -2/3
\end{pmatrix}^{n_2}
\begin{pmatrix}
1 \\ 1 \\ 0
\end{pmatrix}^{n_3}
\begin{pmatrix}
1 \\ 0 \\ 2/3
\end{pmatrix}^{n_4}
\begin{pmatrix}
1 \\ 0 \\ -2/3
\end{pmatrix}^{n_5}
\begin{pmatrix}
1 \\ 0 \\ 0
\end{pmatrix}^{n_6} \nolabel \\
& &\begin{pmatrix}
0 \\ 1 \\ 2/3
\end{pmatrix}^{n_7}
\begin{pmatrix}
0 \\ 1 \\ -2/3
\end{pmatrix}^{n_8}
\begin{pmatrix}
0 \\ 1 \\ 0
\end{pmatrix}^{n_9}
\begin{pmatrix}
0 \\ 0 \\ 2/3
\end{pmatrix}^{n_{10}}
\begin{pmatrix}
0 \\ 0 \\ -2/3
\end{pmatrix}^{n_{11}}. \nolabel 
\end{eqnarray}
We then impose (\ref{eq:squaredconstraint}) on $n_{11}$, $n_9$, and $n_6$, giving 
\begin{eqnarray}
n_{11} &\in& 6\mathbb{Z} - ( n_1 + n_2+n_4+ n_5+n_7+n_8+n_{10}) \nolabel \\
n_9 &\in& 4\mathbb{Z} - (n_1+n_2+n_3+n_7+n_8) \nolabel \\
n_6 &\in& 4\mathbb{Z} - (n_1+n_2+n_3+n_4+n_5).
\end{eqnarray}
So, we have (for example)
\begin{eqnarray}
2 (\vec \alpha^1)^2 &=& 2(n_1+n_2+n_3+n_4+n_5+n_6) \nolabel \\
&=& 2(n_1+n_2+n_3+n_4+n_5 + 4\mathbb{Z} - (n_1+n_2+n_3+n_4+n_5)) \nolabel \\
&\in& 8 \mathbb{Z} = 0 \; \mod \; 8.
\end{eqnarray}
Similarly,
\begin{eqnarray}
3(\vec \alpha^2)^2 &=& 3\cdot {4 \over 9} (n_1+n_2+n_4+n_5+n_7+n_8+n_{10}+n_{11}) \nolabel \\
&=&3\cdot {4 \over 9} (n_1+n_2+n_4+n_5+n_7+n_8+n_{10}+6 \mathbb{Z} \nolabel \\
& & - (n_1+n_2+n_4+n_5+n_7+n_8+n_{10})) \nolabel \\
&\in& 8 \mathbb{Z} = 0 \; \mod \; 8.
\end{eqnarray}
Next, we impose (\ref{eq:dotprodcon}) on $n_8$, $n_5$, and $n_3$.  
\begin{eqnarray}
n_4 \in \mathbb{Z} - (n_1-n_2).
\end{eqnarray}
Thus, 
\begin{eqnarray}
6\vec \alpha^1 \cdot \vec \alpha^3 &=& 6\bigg({2 \over 3}n_1 - {2\over 3}n_2 + {2\over 3}n_4\bigg) \nolabel \\
&=&4(n_1-n_2+(\mathbb{Z}-(n_1-n_2)) \nolabel \\
&\in& 4 \mathbb{Z} = 0 \; \mod \; 4.
\end{eqnarray}
The constraints on $n_8$ and $n_5$ are similar.  

Imposing Modular Invariance in this manner allows us to effectively generate acceptable input sets directly, sparing us the computational difficulty of checking huge numbers of unacceptable sets to find the ones we want.  
So, by designing an algorithm which expresses the sets of $\vec \alpha$'s in terms of $n_i$'s, and imposing (\ref{eq:squaredconstraint}) and (\ref{eq:dotprodcon}), we can systematically generate comprehensive input sets at arbitrary Order and Layer, with a great deal of the work being done up front (imposing (\ref{eq:squaredconstraint}) and (\ref{eq:dotprodcon})).  It remains to discuss how these sets are to be analyzed to generate statistics (including consideration of models with initially matching gauge groups, that may or may not lead to possibilities for different models once matter sectors are added). This we leave for discussion in our upcoming paper.

Before moving on, we make one final observation.  By combining (\ref{eq:new}), (\ref{eq:squaredconstraint}), and (\ref{eq:dotprodcon}), we can write down the approximate number of of sets which satisfy Modular Invariance for a given $L$, $\{N_i\},i=1,\ldots,L$, and $A\equiv \prod_{i=1}^L(N_i)-1$, 
\begin{eqnarray}
\bigg(\prod_{i=1}^L \prod_{j=1}^{i-1} N_{ij}\bigg){\Gamma(23+A)\over 2^LA!(A+1)^L\Gamma(23)}. \label{eq:approx}
\end{eqnarray}
This expression clearly increases monotonically for increasing Layer and Order, implying an infinite number of valid sets.  However, it is likely that past a certain level of complexity the models will become redundant, providing an effective upper limit.  

As an extremely elementary example of how this works, we consider the simplest possible example, $L1O2$.  There is only one $n_i$, and (\ref{eq:squaredconstraint}) simplifies to $n_1 \in 4\mathbb{Z}$, and equation (\ref{eq:approx}) gives $5.75$, which is a reasonable estimation of the obvious $5$ solutions, $n_1 \in \{4,8,12,16,20\}$ (we exclude $n_1 = 0$ for obvious reasons).  Each of these models can be easily analyzed, giving Table 1.

\begin{table} [h]
\caption{The complete set of gauge group models of L1O2 class.}
\centering
\begin{tabular} {|c || c |}
\hline
$n_1$  & Gauge Group \\ [0.5ex]
\hline \hline
4 &   $SO(44)$ \\
\hline
8 & $SO(28)\times E_8$ \\
\hline
12 & $SO(20)\times SO(24)$ \\ 
\hline
16 & $SO(12)\times SO(32)$ \\
\hline
20 & $SO(40)\times SU(2) \times SU(2)$ \\
\hline
\end{tabular}
\label{table:nonlin}
\end{table}


Using this approach, statistics at very high Layer and Order can be produced fairly easily.  The results of such an analysis will be discussed in much more detail in an upcoming paper \cite{robinson2}. We also wish to emphasize that the approach we have introduced here can be generalized to include matter sectors as well, as will be demonstrated in \cite{greenwald1}.

\section{Solving for Alpha}

As a final comment, we can rewrite (\ref{eq:gso}) as 
\begin{eqnarray}
\vec \alpha^i \cdot \vec Q_{\vec V^k} = \sum_{n=1}^Lm^k_nk_{in} + 2a_{ik},
\end{eqnarray}
for some set $a_{ij}$.  

Then, using (\ref{eq:kijone}) and (\ref{eq:kijtwo}), we can write the Cartan Matrix of a given model as 
\begin{eqnarray}
C^{jk} = \vec Q_{\vec V^j} \cdot \vec Q_{\vec V^k} = (k_1^R + 3)m^j_1m^k_1+\sum_{n=2}^L k^R_nm^j_nm^k_n+\phi_{jk}+(\vec m^j)^T \cdot \vec a^k+(\vec m^k)^T\cdot \vec a^j,
\end{eqnarray}
where the vector $\vec k^R$ is a vector with components equal to the diagonal elements of $k_{ij}$, $\vec a^k$ is the vector with $i^{th}$ element $a_{ik}$, and $\phi_{jk} = \vec F^j \cdot \vec F^k$.  Using this, the left hand side can be defined for a specific gauge group, and then the right hand side can be solved as a system of coupled quadratics over positive integers.  The interesting thing is that the right hand side has no $\vec \alpha$ dependence.  A solution to this equation for a given $\vec k^R$, $\phi_{jk}$, and $\vec a^k$ will provide a set of constraints for $\vec \alpha$'s which give the gauge group defined by the desired Cartan Matrix.  

\section{Acknowledgements}

Research funding leading to this manuscript was partially provided by Baylor URC grant 0301533BP.

\end{document}